\documentclass[10pt]{article}%
\usepackage{amsfonts}
\usepackage{amssymb}
\usepackage{graphicx}
\usepackage{amsmath}%
\providecommand{\U}[1]{\protect\rule{.1in}{.1in}}
\textheight
22.0cm \textwidth 14.8cm \topmargin -0.5 cm \oddsidemargin 0.6 cm
\begin{document}

\title{Bose-Einstein condensation temperature of finite systems}
\author{Mi Xie\thanks{Email: xiemi@tju.edu.cn}\\{\footnotesize Department of Physics, School of Science, Tianjin University,
Tianjin 300072, P. R. China}}
\date{}
\maketitle

\begin{abstract}
In the studies of the Bose-Einstein condensation of ideal gases in finite
systems, the divergence problem usually arises in the equation of state. In
this paper, we present a technique based on the heat kernel expansion and the
zeta-function regularization to solve the divergence problem, and obtain the
analytical expression of the Bose-Einstein condensation temperature for
general finite systems. The result is represented by the heat kernel
coefficients, in which the asymptotic energy spectrum of the system is used.
Besides the general case, for the systems with exact spectra, e.g., ideal
gases in an infinite slab or in a three-sphere, the sums of the spectra can be
performed exactly and the calculation of the corrections to the critical
temperatures is more direct. For the system confined in a bounded potential,
the form of the heat kernel is different from the usual heat kernel expansion.
We show that, as long as the asymptotic form of the global heat kernel can be
found out, our method also works. For Bose gases confined in three- and
two-dimensional isotropic harmonic potentials, we obtain the higher-order
corrections to the usual results of the critical temperatures. Our method also
can be applied to the problem of the generalized condensation, and we give the
correction of the boundary on the second critical temperature in a highly
anisotropic slab.

\end{abstract}


\section{Introduction}

After the experimental realization of Bose-Einstein condensation (BEC) in
ultracold atoms \cite{BEC,BEC2,BEC3}, numerous theoretical studies have been
devoted to the BEC phase transition in finite systems. Strictly speaking, a
phase transition can only occur in the infinite system. In a finite system,
all thermodynamic quantities are analytical functions of the temperature, so
there does not exist a genuine phase transition. On the other hand, the
behavior of a large system is practically the same as the infinite one just as
observed in the experiments, so a (quasi-)critical temperature is needed to
discuss the property of the finite system. To define the critical temperature
of BEC in a finite system, many schemes with different criteria have been
discussed, such as a small value of the condensate fraction
\cite{KD,Pathria,KT,Noronha}, the maximal fluctuation or the inflexion point
of the ground-state occupation number \cite{IR}, the maximum of the specific
heat \cite{KT2,HHR}, etc.

To discuss the BEC phase transition, the heat kernel approach
\cite{KirstenBK,Vassilevich,Gilkey} is a powerful tool. In the past years, the
heat kernel approach has been widely applied in many fields of physics,
including quantum field theory \cite{QFT,QFT2,QFT3}, quantum gravity
\cite{gravity,gravity2}, string theory \cite{AGM}, and quantum statistical
mechanics \cite{KirstenBK,PLA2003}. However, when applied to the BEC of an
ideal gas in a finite system, the heat kernel expansion encounters the problem
of divergence. In fact, the problem of divergence is also inevitable in other
approaches besides the heat kernel expansion. If one regards that the phase
transition occurs at the chemical potential $\mu=0$ just as that in the
thermodynamic limit, the equation of state of the gas will be divergent
\cite{Pathria,Noronha,Biswas}. To avoid this difficulty, the discussion of the
phase transition in finite systems usually ignores the divergent terms
\cite{KD,KT2,HHR,KirstenBK,FK,GH2}, or, for the systems with exact energy
spectra, directly performs the sum of the spectra \cite{Pathria,NT,Cheng}.

In this paper, we will present a technique to solve the divergence problem in
the BEC phase transition of nonrelativistic ideal gases confined in finite
systems, with the help of the heat kernel expansion and the zeta-function
regularization. By using the asymptotic spectrum calculated from the heat
kernel expansion, we obtain the analytical expression of the critical
temperature for ideal Bose gases in general finite systems, which is expressed
only by the heat kernel coefficients. In this result, the effect of the finite
number of particles is separated from other factors. We compare it with the
numerical result of the specific heat of an ideal Bose gas in a cube with
period boundary conditions, and they fit very well. Our result also give the
influence of the boundary on the critical temperature for an arbitrary cavity.
Furthermore, when the sum of the energy spectrum can be performed exactly, we
can replace the asymptotic result by the exact sum and obtain a more precise
critical temperature. The critical temperatures of BEC in an infinite slab and
in a three-sphere $S^{3}$ are given as examples, and the results are
consistent with those calculated by other methods. In a bounded potential, the
heat kernel has a different form from the usual heat kernel expansion. Our
method also can be applied to such cases. As examples, we give the
higher-order corrections to the critical temperatures of BEC in three- and
two-dimensional isotropic harmonic potentials. In addition, in highly
anisotropic systems the condensation may not occur in the ground state but
distributes in a set of single-particle states, i.e., the generalized BEC
\cite{BL,BLP,BLL}. The Bose gases in such systems may undergo two kinds of
phase transitions. We will also consider the influence of the boundary on the
second critical temperature in a highly anisotropic slab.

The paper is organized as follows. In section \ref{II}, we discuss the BEC
phase transition of an ideal gas in a general finite system and express the
critical temperature analytically in terms of the heat kernel coefficients. As
examples, we discuss the influences of the finite number of particles and the
boundary, respectively. In section \ref{III}, we calculate the critical
temperatures of BEC in an infinite slab and in a three-sphere. In these cases
the sums of the energy spectra are performed exactly. In section \ref{IV}, we
consider the phase transition in three- and two-dimensional isotropic harmonic
traps, and obtain the higher-order corrections to the critical temperatures.
In section \ref{V}, we discuss the correction of the boundary on the second
critical temperature in a highly anisotropic slab. The conclusion and some
discussions are presented in section \ref{VI}.

\section{Critical temperature of BEC in finite systems \label{II}}

In the grand canonical ensemble, the grand potential of an ideal Bose gas is%
\begin{equation}
\ln\Xi=-\sum_{i}\ln\left(  1-ze^{-\beta E_{i}}\right)  ,
\end{equation}
where $\left\{  E_{i}\right\}  $ is the single-particle energy spectrum, the
fugacity $z=e^{\beta\mu}$ with $\mu$ the chemical potential, and
$\beta=1/\left(  k_{B}T\right)  $ with $k_{B}$ the Boltzmann constant.
Expanding the logarithmic term gives%
\begin{equation}
\ln\Xi=\sum_{\ell=1}^{\infty}\frac{1}{\ell}z^{\ell}\sum_{i}e^{-\ell\beta
E_{i}}.\label{Xi}%
\end{equation}
Since the energy spectrum satisfy the eigenvalue equation%
\begin{equation}
\left[  -\nabla^{2}+\frac{2m}{\hbar^{2}}V\left(  \mathbf{x}\right)  \right]
\psi_{i}=\frac{2m}{\hbar^{2}}E_{i}\psi_{i},
\end{equation}
the sum over the spectrum in eq. (\ref{Xi}) can be expressed as the global
heat kernel of the operator $D=-\nabla^{2}+\left(  2m/\hbar^{2}\right)
V\left(  \mathbf{x}\right)  $. Mathematically, the global heat kernel of an
operator $D$ is defined as%
\begin{equation}
K\left(  t\right)  =\sum_{i}e^{-\lambda_{i}t},\label{Kt}%
\end{equation}
where $\left\{  \lambda_{i}\right\}  $ is the spectrum of the operator $D$.
When $t\rightarrow0$, it can be asymptotically expanded as a series of $t$,
which is the heat kernel expansion,
\begin{equation}
K\left(  t\right)  \approx\frac{1}{\left(  4\pi t\right)  ^{3/2}}%
\sum_{k=0,\frac{1}{2},1,\cdots}^{\infty}B_{k}t^{k},~~\left(  t\rightarrow
0\right) \label{HKexp}%
\end{equation}
where $B_{k}$ $\left(  k=0,1/2,1,\cdots\right)  $ are the heat kernel
coefficients. Therefore, with the help of the heat kernel expansion, the grand
potential in eq. (\ref{Xi}) can be rewritten as%

\begin{equation}
\ln\Xi=\sum_{\ell=1}^{\infty}\frac{1}{\ell}K\left(  \ell\frac{\hbar^{2}\beta
}{2m}\right)  z^{\ell}=\frac{1}{\lambda^{3}}\sum_{k=0,\frac{1}{2},1,\cdots
}^{\infty}\frac{B_{k}}{\left(  4\pi\right)  ^{k}}\lambda^{2k}g_{5/2-k}\left(
z\right)  ,\label{qexp}%
\end{equation}
where%
\begin{equation}
g_{\sigma}\left(  z\right)  =\frac{1}{\Gamma(\sigma)}\int_{0}^{\infty}%
\frac{x^{\sigma-1}}{z^{-1}e^{x}-1}dx=\sum_{k=1}^{\infty}\frac{z^{k}}%
{k^{\sigma}}%
\end{equation}
is the Bose-Einstein integral, and $\lambda=\sqrt{2\pi\beta}\hbar/\sqrt{m}$ is
the mean thermal wavelength. For a system with the fixed number of particles,
the relation between the fugacity $z$ and the temperature $T$ is given by the
equation of the particle number%
\begin{equation}
N=\left(  z\frac{\partial\ln\Xi}{\partial z}\right)  _{V,T}=\frac{1}%
{\lambda^{3}}\sum_{k=0,\frac{1}{2},1,\cdots}^{\infty}\frac{B_{k}}{\left(
4\pi\right)  ^{k}}\lambda^{2k}g_{3/2-k}\left(  z\right)  ,\label{Nexp}%
\end{equation}
or,%
\begin{equation}
n\lambda^{3}=g_{3/2}\left(  z\right)  +\frac{B_{1/2}}{\sqrt{4\pi}V}\lambda
g_{1}\left(  z\right)  +\frac{B_{1}}{4\pi V}\lambda^{2}g_{1/2}\left(
z\right)  +\cdots,\label{nlambda}%
\end{equation}
where $n=N/V$ is the particle number density, and we have taken $B_{0}=V$.

Clearly, the first terms in eqs. (\ref{qexp}) and (\ref{nlambda}) give the
ordinary equation of state for an ideal Bose gas in the thermodynamic limit.
The other terms in these equations describe the difference between the finite
system and that in the thermodynamic limit. To discuss the BEC phase
transition in the thermodynamic limit, one only needs to set $z=1$ or the
chemical potential $\mu=0$ in the equation of the particle number, then the
critical temperature can be obtained immediately. However, since the
Bose-Einstein integral $g_{\sigma}\left(  1\right)  $ is divergent at
$\sigma\leq1$, the correction terms in eq. (\ref{nlambda}) become singular at
$\mu=0$. In other words, the traditional method for determining the critical
temperature is invalid for the system described by eq. (\ref{nlambda}).

In the following, we will show that, based on the heat kernel expansion and
the zeta-function regularization, we can solve the divergence problem and
obtain an analytical expression of the critical temperature of BEC.

At the transition point, the chemical potential $\mu\rightarrow0$, and the
leading term of the asymptotic expression of the Bose-Einstein integral is%
\begin{equation}
g_{\sigma}\left(  e^{\beta\mu}\right)  \approx\left\{
\begin{array}
[c]{lll}%
\zeta\left(  \sigma\right)  , & \left(  \sigma\geq\frac{3}{2}\right)  & \\
-\ln\left(  -\beta\mu\right)  , & \left(  \sigma=1\right)  & \\
\Gamma\left(  -\sigma+1\right)  \frac{1}{\left(  -\beta\mu\right)
^{-\sigma+1}}, & \left(  \sigma\leq\frac{1}{2}\right)  & ~~~\left(
\mu\rightarrow0\right)
\end{array}
\right. \label{BEint}%
\end{equation}
where $\zeta\left(  \sigma\right)  =\sum_{n=1}^{\infty}n^{-\sigma}$ is the
Riemann zeta function. Expanding eq. (\ref{nlambda}) and keeping the leading
term of the Bose-Einstein integrals, we reach its asymptotic expression for
$\mu\rightarrow0$ as
\begin{equation}
n\lambda^{3}\approx\zeta\left(  \frac{3}{2}\right)  -\frac{B_{1/2}}{\sqrt
{4\pi}V}\lambda\ln\left(  -\beta\mu\right)  +I,\label{nlambda3}%
\end{equation}
where
\begin{equation}
I=\sum_{k=1,\frac{3}{2},2,\cdots}^{\infty}\frac{B_{k}}{\left(  4\pi\right)
^{k}V}\lambda^{2k}\Gamma\left(  k-\frac{1}{2}\right)  \frac{1}{\left(
-\beta\mu\right)  ^{k-1/2}}\label{I0}%
\end{equation}
is introduced for simplification. When $\mu\rightarrow0$, this is a series in
which every term is divergent. However, it can be summed up by use of the heat
kernel expansion and eq. (\ref{nlambda3}) will give an analytical result. We
will show the procedure in the following.

First, we represent the gamma function in eq. (\ref{I0}) as an integral and
introduce a regularization parameter $s$ which will be set to $0$ to deal with
the divergence problem, i.e.,%
\begin{equation}
\Gamma\left(  \xi\right)  =\int_{0}^{\infty}x^{\xi-1+s}e^{-x}dx.~~~\left(
s\rightarrow0\right) \label{Gamma}%
\end{equation}
Then eq. (\ref{I0}) becomes%

\begin{equation}
I=\frac{\sqrt{-\beta\mu}}{V}\int_{0}^{\infty}dxx^{s-3/2}e^{-x}\sum
_{k=1,\frac{3}{2},2,\cdots}^{\infty}B_{k}\left(  \frac{\hbar^{2}}{2m}\frac
{x}{-\mu}\right)  ^{k}.\label{I1}%
\end{equation}
The sum in the integral differs from the heat kernel expansion eq.
(\ref{HKexp}) just by two extra terms and a common coefficient, so we can
express eq. (\ref{I1}) by the global heat kernel as%
\begin{align}
I  &  =\frac{\sqrt{-\beta\mu}}{V}\int_{0}^{\infty}dxx^{s-3/2}e^{-x}\left[
\left(  4\pi\frac{\hbar^{2}}{2m}\frac{x}{-\mu}\right)  ^{3/2}K\left(
\frac{\hbar^{2}}{2m}\frac{x}{-\mu}\right)  -B_{0}-B_{1/2}\left(  \frac
{\hbar^{2}}{2m}\frac{x}{-\mu}\right)  ^{1/2}\right] \nonumber\\
&  =\frac{\lambda^{3}}{V}\frac{1}{-\beta\mu}\int_{0}^{\infty}dxx^{s}%
e^{-x}K\left(  \frac{\hbar^{2}}{2m}\frac{x}{-\mu}\right)  -\Gamma\left(
s-\frac{1}{2}\right)  \sqrt{-\beta\mu}-\Gamma\left(  s\right)  \frac
{B_{1/2}\lambda}{\sqrt{4\pi}V}.\label{I2}%
\end{align}
As a result, the divergent sum in eq. (\ref{I0}) is converted to the global
heat kernel.

Next, the integral in the first term in eq. (\ref{I2}) can be performed by
substituting the definition of the global heat kernel eq. (\ref{Kt}),%
\begin{equation}
\int_{0}^{\infty}dxx^{s}e^{-x}K\left(  \frac{\hbar^{2}}{2m}\frac{x}{-\mu
}\right)  =\sum_{i=1}^{\infty}\int_{0}^{\infty}dxx^{s}e^{-x}e^{-E_{i}\frac
{x}{-\mu}}=\Gamma\left(  1+s\right)  \sum_{i=1}^{\infty}\frac{\left(
-\mu\right)  ^{1+s}}{\left(  E_{i}-\mu\right)  ^{1+s}}.\label{HK2sum}%
\end{equation}
Our aim is to determine the critical temperature of the BEC, which occurs at
$\mu\rightarrow0$. At this limit, eq. (\ref{I2}) becomes%
\begin{equation}
I=\Gamma\left(  1+s\right)  \frac{\lambda^{3}}{V}\frac{\left(  -\mu\right)
^{s}}{\beta}\sum_{i=1}^{\infty}\frac{1}{E_{i}^{1+s}}-\Gamma\left(  s\right)
\frac{B_{1/2}\lambda}{\sqrt{4\pi}V}.\label{I3}%
\end{equation}
Note that the sum of the spectrum is indeed the spectrum zeta function defined
as $\zeta_{s}\left(  \sigma\right)  =\sum_{i=1}^{\infty}\lambda_{i}^{-\sigma}%
$, which gives%

\begin{equation}
\sum_{i=1}^{\infty}\frac{1}{E_{i}^{1+s}}=\left(  \frac{2m}{\hbar^{2}}\right)
^{1+s}\sum_{i=1}^{\infty}\frac{1}{\lambda_{i}^{1+s}}=\left(  \frac{2m}%
{\hbar^{2}}\right)  ^{1+s}\zeta_{s}\left(  1+s\right)  .\label{s-zeta}%
\end{equation}
Eq. (\ref{nlambda3}) now is%
\begin{equation}
n\lambda^{3}=\zeta\left(  \frac{3}{2}\right)  -\frac{B_{1/2}\lambda}%
{\sqrt{4\pi}V}\ln\left(  -\beta\mu\right)  +\Gamma\left(  1+s\right)
\frac{\lambda^{3}}{V}\frac{\left(  -\mu\right)  ^{s}}{\beta}\sum_{i=1}%
^{\infty}\frac{1}{E_{i}^{1+s}}-\Gamma\left(  s\right)  \frac{B_{1/2}\lambda
}{\sqrt{4\pi}V}.\label{nlambdaf}%
\end{equation}
From this result we can find that although it is based on the whole heat
kernel expand, only the first two heat kernel coefficients $B_{0}=V$ and
$B_{1/2}$ appear in this expression. However, it does not mean that the
higher-order heat kernel coefficients are irrelevant to the critical
temperature. In eq. (\ref{nlambdaf}), there is a term of the sum of the
spectrum, and the information of the spectrum is embodied in the heat kernel coefficients.

Finally, we will deal with the sum in eq. (\ref{nlambdaf}). For a general
system, the exact spectrum is not known, but we can obtain its asymptotic
expression on the basis of the heat kernel expansion. As given in Ref.
\cite{JHEP09}, we can first obtain the counting function from the heat kernel
expansion, then achieve the asymptotic expansion of the spectrum from the
counting function. Specifically, the counting function $N\left(  \chi\right)
$ is defined as the number of the eigenstates of an operator with the
eigenvalue smaller than $\chi$. The relation between the counting function and
the global heat kernel is \cite{JHEP09}%
\begin{equation}
N\left(  \chi\right)  =\frac{1}{2\pi i}\int_{c-i\infty}^{c+i\infty}K\left(
t\right)  \frac{e^{\chi t}}{t}dt.\label{N-K}%
\end{equation}
From the heat kernel expansion eq. (\ref{HKexp}), we can calculate the
asymptotic expansion of the counting function%
\begin{equation}
N\left(  \chi\right)  =\frac{1}{\left(  4\pi\right)  ^{3/2}}\sum_{k=0,\frac
{1}{2},1,\cdots}^{\infty}\frac{B_{k}}{\Gamma\left(  5/2-k\right)  }%
\chi^{3/2-k}.
\end{equation}
Setting%
\begin{equation}
N\left(  \lambda_{i}\right)  =i
\end{equation}
will give the asymptotic expansion of the spectrum. The first two terms read
\cite{JHEP09}%
\begin{equation}
\lambda_{i}\approx\left(  6\pi^{2}\right)  ^{2/3}\frac{1}{B_{0}^{2/3}}%
i^{2/3}-\left(  \frac{3}{4}\right)  ^{1/3}\pi^{7/6}\frac{B_{1/2}}{B_{0}^{4/3}%
}i^{1/3}.\label{Lambda_n}%
\end{equation}
Thus the spectrum of a general system is represented by the heat kernel
coefficients asymptotically.

Since the energy spectrum $E_{i}=\left(  \hbar^{2}/2m\right)  \lambda_{i}$,
from eq. (\ref{Lambda_n}) we have%
\begin{align}
&  \sum_{i=1}^{\infty}\frac{1}{E_{i}^{1+s}}\approx\sum_{i=1}^{\infty}\left[
\frac{2m}{\hbar^{2}}\left(  \frac{B_{0}}{6\pi^{2}}\right)  ^{2/3}\right]
^{1+s}\left[  \frac{1}{i^{2/3+2s/3}}+\frac{\left(  1+s\right)  B_{1/2}}%
{2\cdot6^{1/3}\pi^{1/6}B_{0}^{2/3}}\frac{1}{i^{1+2s/3}}\right] \nonumber\\
&  =\left[  \frac{2m}{\hbar^{2}}\left(  \frac{V}{6\pi^{2}}\right)
^{2/3}\right]  ^{1+s}\left[  \zeta\left(  \frac{2}{3}+\frac{2s}{3}\right)
+\frac{\left(  1+s\right)  B_{1/2}}{2\cdot6^{1/3}\pi^{1/6}V^{2/3}}\zeta\left(
1+\frac{2s}{3}\right)  \right]  .\label{sum_En}%
\end{align}
In this calculation, we have represented the sum by the Riemann zeta function.
Note that by the zeta-function regularization, the two divergent sums are
replace by zeta functions. However, the second term in eq. (\ref{sum_En})
still contains a function $\zeta\left(  1+\frac{2s}{3}\right)  $ which is
divergent at $s\rightarrow0$. Substituting this result into eq. (\ref{I3}) and
expanding it around $s=0$ gives%

\begin{equation}
I\approx\frac{4\zeta\left(  2/3\right)  }{6^{2/3}\pi^{1/3}}\frac{\lambda
}{V^{1/3}}-\frac{B_{1/2}\lambda}{\sqrt{4\pi}V}\left(  \ln\frac{3^{2/3}%
\pi^{4/3}\hbar^{2}}{2^{1/3}mV^{2/3}\left(  -\mu\right)  }-1-\frac{2}{3}%
\gamma_{E}\right)  ,\label{I4}%
\end{equation}
where $\gamma_{E}=0.5772$ is the Euler constant. We find that the linearly
divergent terms of $s$ from the zeta function and the gamma function have been
canceled. On the other hand, there is a logarithmically divergent term of
$\mu$ in this equation, but this term also will be canceled when substituting
eq. (\ref{I4}) into eq. (\ref{nlambda3}):%
\begin{equation}
n\lambda^{3}=\zeta\left(  \frac{3}{2}\right)  +\frac{4\zeta\left(  2/3\right)
}{6^{2/3}\pi^{1/3}}\frac{\lambda}{V^{1/3}}-\frac{B_{1/2}\lambda}{\sqrt{4\pi}%
V}\left(  \ln\frac{3^{2/3}\pi^{1/3}\lambda^{2}}{2^{4/3}V^{2/3}}-1-\frac{2}%
{3}\gamma_{E}\right)  .\label{PT}%
\end{equation}

By using the above technique, our final result eq. (\ref{PT}) is fully
analytical now. This equation holds at the transition point, so the critical
temperature can be solved straightforward. Clearly, the first term in the
right-hand-side of eq. (\ref{PT}) gives the critical temperature in the
thermodynamic limit%

\begin{equation}
T_{0}=\frac{2\pi\hbar^{2}n^{2/3}}{\zeta^{2/3}\left(  3/2\right)  mk_{B}}.
\end{equation}
Regarding the last two terms in eq. (\ref{PT}) as small corrections, we can
obtain the critical temperature $T_{c}$, satisfying%
\begin{align}
\frac{T_{c}-T_{0}}{T_{0}}  &  \approx\frac{2}{9\zeta^{2/3}\left(  3/2\right)
}\left[  -\frac{3^{1/3}4^{2/3}\zeta\left(  2/3\right)  }{\pi^{1/3}}-\frac
{1}{\sqrt{\pi}}\left(  \ln\frac{4N}{3\sqrt{\pi}\zeta\left(  3/2\right)
}+\frac{3}{2}+\gamma_{E}\right)  \frac{B_{1/2}}{V^{2/3}}\right]  \frac
{1}{N^{1/3}}\nonumber\\
&  =\left[  0.7115-\left(  0.06607\ln N+0.05506\allowbreak\right)
\frac{B_{1/2}}{V^{2/3}}\right]  \frac{1}{N^{1/3}}.\label{delta}%
\end{align}

This result shows that the correction to the critical temperature of an ideal
Bose gas in a finite system is represented by the heat kernel coefficients.
Since the typical case of nonvanishing $B_{1/2}$ is the system with a
boundary, one direct application of this result is to describe the influence
of boundary on the critical temperature. Besides, the first term in eq.
(\ref{delta}) is proportional to $1/N^{1/3}$, or, it only related to the
particle number. It indicates that this term describes the pure contribution
of the finite number of particles.

Eq. (\ref{delta}) only contains two terms since we only consider the first two
terms in the asymptotic expression of the spectrum eq. (\ref{Lambda_n}). More
contributions can be easily obtained by including more terms in eq.
(\ref{Lambda_n}), and each of them will contribute a term proportional to some
$\zeta\left(  \alpha\right)  $ with $\alpha>1$, so there is no divergence in
these terms and the calculation is straightforward. It is easy to check that
all these contributions are also proportional to $1/N^{1/3}$.

In the following, we will consider two examples to show the correction in eq.
(\ref{delta}) to the critical temperature.

\subsection{Effect of the finite number of particles}

According to eq. (\ref{delta}), if the only influence on the ideal gas is the
finite number of particles, the critical temperature of the BEC will be
\begin{equation}
\frac{T_{c}-T_{0}}{T_{0}}=-\frac{8\zeta\left(  2/3\right)  }{3\cdot6^{2/3}%
\pi^{1/3}\zeta^{2/3}\left(  3/2\right)  }\frac{1}{N^{1/3}}=0.7115\frac
{1}{N^{1/3}}.\label{Tc_N}%
\end{equation}
Therefore, the critical temperature increases in a finite system compared with
that in the thermodynamic limit.

To check this result, we consider an ideal Bose gas confined in a
three-dimensional cube of side length $L$ with period boundary conditions. The
single-particle energy spectrum is
\begin{equation}
E\left(  n_{x},n_{y},n_{z}\right)  =\frac{2\pi^{2}\hbar^{2}}{mL^{2}}\left(
n_{x}^{2}+n_{y}^{2}+n_{z}^{2}\right)  .~~\left(  n_{x},n_{y},n_{z}=0,\pm
1,\pm2,\cdots\right)
\end{equation}
The corresponding global heat kernel can be obtained by repeatedly applying
the Euler-MacLaurin formula \cite{PRE04}%
\begin{equation}
K\left(  \frac{\hbar^{2}\beta}{2m}\right)  =\sum_{n_{x},n_{y},n_{z}=-\infty
}^{\infty}e^{-\beta E\left(  n_{x},n_{y},n_{z}\right)  }\approx\left(
\frac{2\pi\hbar^{2}\beta}{m}\right)  ^{-3/2}V.\label{K_PBT}%
\end{equation}
The heat kernel expansion only contains one term just like that in infinite
space, but it is not an exact result since some exponentially small terms have
been neglected. This form implies that the only factor affecting the critical
temperature in such a system is the volume, or, the particle number. In other
words, the critical temperature of BEC in this system should satisfy eq.
(\ref{Tc_N}).

\begin{figure}[ptb]
\begin{center}
\includegraphics[height=3in]{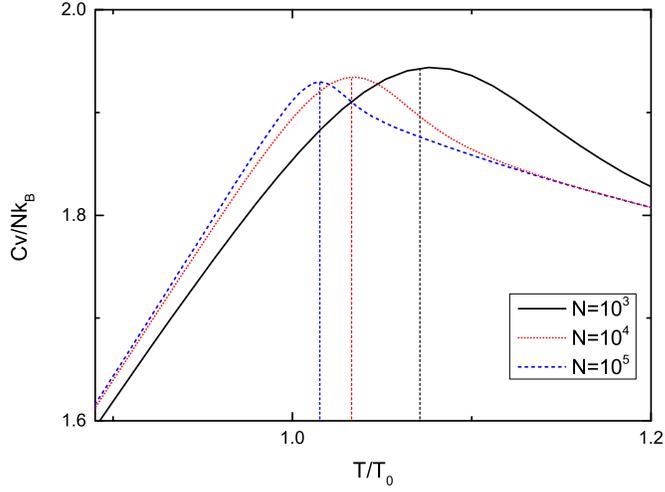}
\end{center}
\caption{The numerical result of the specific heat versus the temperature for
an ideal Bose gas in a cube with period boundary conditions. The three curves
represent the cases of the particle number $N=10^{3}$ (black solid line),
$N=10^{4}$ (red dotted line), and $N=10^{5}$ (blue dashed line), respectively.
The critical temperatures given by eq. (\ref{Tc_N}) for these cases are
represented by the vertical dashed lines.}%
\label{fig1}%
\end{figure}

In figure \ref{fig1} we plot the exact numerical result of the specific heat
versus the temperature for an ideal Bose gas in the cube with period boundary
conditions. The critical temperature given by eq. (\ref{Tc_N}) agrees very
well with the maximum of the specific heat.

\subsection{Effect of the boundary}

The influence of the boundary on BEC with different boundary conditions has
motivated many studies \cite{GH,VVZ,MZ}. In the following, we will discuss the
influence of the boundary on the critical temperature according to eq.
(\ref{delta}). We know that at a manifold with a boundary, the coefficient
$B_{1/2}$ reflects the leading effect of the boundary \cite{Vassilevich}%
\begin{equation}
B_{1/2}=\mp\frac{\sqrt{\pi}}{2}S,\label{B1/2}%
\end{equation}
where $S$ is the surface area of the system, the signs $-$ and $+$ correspond
Dirichlet and Neumann boundary conditions, respectively. Substituting eq.
(\ref{B1/2}) into eq. (\ref{delta}) gives the critical temperature for the
corresponding system.

To check the result, we consider a Bose gas confined in a cube of side length
$L$ with Dirichlet boundary conditions. The energy spectrum is%
\begin{equation}
E\left(  n_{x},n_{y},n_{z}\right)  =\frac{\pi^{2}\hbar^{2}}{2mL^{2}}\left(
n_{x}^{2}+n_{y}^{2}+n_{z}^{2}\right)  ,~~\left(  n_{x},n_{y},n_{z}%
=1,2,3,\cdots\right)
\end{equation}
and the heat kernel coefficients are \cite{JHEP09}%
\begin{align}
B_{0}  &  =L^{3},B_{1/2}=-3\sqrt{\pi}L^{2},B_{1}=3\pi L,\nonumber\\
B_{3/2}  &  =-\pi^{3/2},B_{k}=0\left(  k>\frac{3}{2}\right)  .
\end{align}
Then from eq. (\ref{delta}), the correction to the critical temperature is
\begin{align}
\frac{T_{c}-T_{0}}{T_{0}}  &  =\frac{2}{3\zeta^{2/3}\left(  3/2\right)
}\left[  -\frac{4^{2/3}\zeta\left(  2/3\right)  }{3^{2/3}\pi^{1/3}}+\left(
\ln\frac{4N}{3\sqrt{\pi}\zeta\left(  3/2\right)  }+\frac{3}{2}+\gamma
_{E}\right)  \right]  \frac{1}{N^{1/3}}\nonumber\\
&  =\left(  0.3515\ln N+1.004\right)  \frac{1}{N^{1/3}}.\label{delta_boundary}%
\end{align}
Clearly, this result contains the contributions from the finite number of
particles and the boundary. As mentioned above, in eq. (\ref{delta_boundary})
only the first-order correction of the spectrum is included. If included more
terms in the asymptotic expansion of the spectrum eq. (\ref{Lambda_n}), we can
obtain a more precise critical temperature. The final result converges to%

\begin{equation}
\frac{T_{c}-T_{0}}{T_{0}}=\left(  0.3515\ln N+0.257\allowbreak\right)
\frac{1}{N^{1/3}}.\label{delta_inf_boundary}%
\end{equation}
The $\ln N$ term in the parentheses is consistent with Ref. \cite{GH}.

\section{Systems with exact spectra \label{III}}

In section \ref{II}, we have given a general discussion on the critical
temperature of BEC, in which the asymptotic expansion of the spectrum eq.
(\ref{Lambda_n}) is taken into account. However, if the spectrum of the system
is known and can be summed exactly, we can perform the sum and obtain a more
precise critical temperature. In the following we will consider two examples:
the infinite slab and the three-sphere $S^{3}$.

\subsection{BEC in the infinite slab \label{slab}}

Consider a Bose gas between two infinite parallel planes with distance $L$. It
can be regarded as a rectangular box of side lengths $L_{x}=L_{y}=a$ $\left(
a\rightarrow\infty\right)  $ and $L_{z}=L$ with Dirichlet boundary conditions.
The single-particle energy spectrum is then%
\begin{equation}
E\left(  n_{x},n_{y},n_{z}\right)  =\frac{\pi^{2}\hbar^{2}}{2m}\left(
\frac{n_{x}^{2}}{a^{2}}+\frac{n_{y}^{2}}{a^{2}}+\frac{n_{z}^{2}}{L^{2}%
}\right)  .~~\left(  n_{x},n_{y},n_{z}=1,2,3,\cdots\right) \label{En_slab}%
\end{equation}
For such a form of the spectrum, the sum in eq. (\ref{s-zeta}) can be
performed exactly. Since $a\rightarrow\infty$, the sums of $n_{x}$ and $n_{y}
$ are converted to integrals, we have%
\begin{align}
&  \sum_{n_{x},n_{y},n_{z}=1}^{\infty}\frac{1}{E\left(  n_{x},n_{y}%
,n_{z}\right)  ^{1+s}}=\left(  \frac{\pi^{2}\hbar^{2}}{2m}\right)  ^{-\left(
1+s\right)  }\sum_{n_{z}=1}^{\infty}\int_{0}^{\infty}dn_{x}dn_{y}\frac
{1}{\left(  \frac{n_{x}^{2}}{a^{2}}+\frac{n_{y}^{2}}{a^{2}}+\frac{n_{z}^{2}%
}{L^{2}}\right)  ^{1+s}}\nonumber\\
&  =\left(  \frac{\pi^{2}\hbar^{2}}{2m}\right)  ^{-\left(  1+s\right)  }%
\sum_{n_{z}=1}^{\infty}\frac{\pi a^{2}}{4s}\left(  \frac{n_{z}^{2}}{L^{2}%
}\right)  ^{-s}=\left(  \frac{\pi^{2}\hbar^{2}}{2m}\right)  ^{-\left(
1+s\right)  }\frac{\pi a^{2}L^{2s}}{4}\zeta\left(  2s\right)  \frac{1}{s}.
\end{align}
The heat kernel coefficients for the Laplace operator can also be calculated
from the spectrum eq. (\ref{En_slab}),%
\begin{equation}
B_{0}=V=a^{2}L,B_{1/2}=-\sqrt{\pi}a^{2},B_{k}=0\left(  k>\frac{1}{2}\right)  .
\end{equation}
At the transition point, $\mu\rightarrow0$, eq. (\ref{nlambdaf}) becomes%
\begin{equation}
n\lambda^{3}=\zeta\left(  \frac{3}{2}\right)  +\frac{\lambda}{2L}\ln\left(
-\beta\mu\right)  +\frac{\lambda}{L^{1-2s}}\left(  -\frac{2m\mu}{\pi^{2}%
\hbar^{2}}\right)  ^{s}\Gamma\left(  1+s\right)  \zeta\left(  2s\right)
\frac{1}{s}+\frac{\lambda}{2L}\Gamma\left(  s\right)  .
\end{equation}
When $s\rightarrow0$, all of the divergent terms in this expression are
canceled, we have
\begin{equation}
n\lambda^{3}=\zeta\left(  \frac{3}{2}\right)  +\frac{\lambda}{2L}\ln\left(
\frac{\lambda^{2}}{16\pi L^{2}}\right)  .\label{n_slab}%
\end{equation}
This equation holds at the transition point. The second term represents the
correction from the boundary. Then we can obtain the correction to the
critical temperature%
\begin{align}
\frac{T_{c}-T_{0}}{T_{0}}  &  \approx\frac{1}{3\zeta^{2/3}\left(  3/2\right)
}\ln\left(  \frac{16\pi}{\zeta^{2/3}\left(  3/2\right)  }n^{2/3}L^{2}\right)
\frac{1}{n^{1/3}L}\nonumber\\
&  =\left[  0.3514\ln\left(  n^{1/3}L\right)  +0.5758\right]  \frac{1}%
{n^{1/3}L}.
\end{align}

This result is consistent with that in Ref. \cite{NT}, in which the critical
temperature is obtained by use of the Mellin-Barnes transform.

\subsection{BEC in the three-sphere}

The spectrum of the Laplace operator in a three-sphere $S^{3}$ of radius $R$
is \cite{Lachieze-Rey}%

\begin{equation}
\lambda_{n}=n\left(  n+2\right)  \frac{1}{R^{2}}\label{spectrum_S3}%
\end{equation}
with the degeneracy $\left(  n+1\right)  ^{2}$. The global heat kernel is
\cite{Nagase}%
\begin{equation}
K\left(  t\right)  =\frac{V}{\left(  4\pi t\right)  ^{3/2}}e^{\frac{t}{R^{2}}%
},
\end{equation}
where $V=2\pi^{2}R^{3}$ is the volume of $S^{3}$. Therefore, the heat kernel
expansion in $S^{3}$ does not contain half-integer powers of $t$. In fact,
this result can be obtained directly without the specific form of the heat
kernel: Since $S^{3}$ is a smooth manifold without boundary, the heat kernel
expansion will not contain half-integer power terms.

According to the analysis in section \ref{II}, the critical temperature
satisfies eq. (\ref{nlambdaf}). For the $S^{3}$ case, the coefficient
$B_{1/2}=0$, which makes the divergent terms in the general case vanish, so we
can take $s=0$ in the expression, i.e.,%
\begin{equation}
n\lambda^{3}=\zeta\left(  \frac{3}{2}\right)  +\frac{\lambda^{3}}{V}\frac
{1}{\beta}\sum_{i=1}^{\infty}\frac{1}{E_{i}}.\label{nlambda_S3}%
\end{equation}
Then we perform the sum by using the exact spectrum eq. (\ref{spectrum_S3}).
The sum of all the excited states is%
\begin{align}
\sum_{i=1}^{\infty}\frac{1}{E_{i}}  &  =\frac{2mR^{2}}{\hbar^{2}}\sum
_{n=1}^{\infty}\frac{\left(  n+1\right)  ^{2}}{n\left(  n+2\right)  }%
=\frac{2mR^{2}}{\hbar^{2}}\sum_{n=2}^{\infty}\frac{n^{2}}{n^{2}-1}\nonumber\\
&  =\frac{2mR^{2}}{\hbar^{2}}\sum_{n=2}^{\infty}\sum_{k=0}^{\infty}%
n^{-2k}=\frac{2mR^{2}}{\hbar^{2}}\sum_{k=0}^{\infty}\left(  \zeta\left(
2k\right)  -1\right) \nonumber\\
&  =-\frac{3mR^{2}}{2\hbar^{2}},
\end{align}
where we have used eq. (215) in Ref. \cite{Srivastava}%
\begin{equation}
\sum_{k=1}^{\infty}\left(  \zeta\left(  2k\right)  -1\right)  =\frac{3}{4}.
\end{equation}
Then eq. (\ref{nlambda_S3}) becomes%
\begin{equation}
n\lambda^{3}=\zeta\left(  \frac{3}{2}\right)  -\frac{3}{2^{2/3}\pi^{1/3}}%
\frac{\lambda}{V^{1/3}},
\end{equation}
and the correction to the critical temperature is%
\begin{equation}
\frac{T_{c}-T_{0}}{T_{0}}\approx\frac{2^{1/3}}{\pi^{1/3}\zeta^{2/3}\left(
\frac{3}{2}\right)  }\frac{1}{N^{1/3}}=0.4535\frac{1}{N^{1/3}}.\label{Tc_S3}%
\end{equation}

In Ref. \cite{NT}, the authors have also discussed the BEC in $S^{3}$. Their
result has a different coefficient from eq. (\ref{Tc_S3}), but the reason is
that the spectrum in Ref. \cite{NT} is approximately taken as $\lambda
_{n}=\left(  n+1\right)  ^{2}/R^{2}$. If we use this spectrum too, our result
will be the same as that in Ref. \cite{NT}.

\section{BEC in harmonic potentials \label{IV}}

In section \ref{II}, our discussion is based on the heat kernel expansion eq.
(\ref{HKexp}). On the other hand, if an ideal Bose gas is confined in a
bounded potential, the form of the heat kernel will change. In this section,
we will show that even if the form of the heat kernel is different from eq.
(\ref{HKexp}), the above method can still be used to determine the critical
temperature of BEC. We will take three- and two-dimensional harmonic
potentials as examples.

\subsection{Three-dimensional harmonic potentials}

In the literature, many studies are devoted to the BEC phase transition in
three-dimensional ($3D$) harmonic potentials since the realization of the BEC
of ultracold atoms is in such potentials. In a harmonic potential, the heat
kernel of the Laplace operator has a different form from eq. (\ref{HKexp}),
but it still can be expanded as a series. At the transition point, one also
will encounter the problem of divergence. In many theoretical studies on the
critical temperature of BEC in harmonic potentials based on the heat kernel
approach, the divergent terms are ignored \cite{KD,KT2,HHR,GH2}. In this
section, we will apply the technique provided in section \ref{II} to consider
all the terms in the equation of state. By solving the divergence problem, we
will give the analytical result of the critical temperature with the
higher-order correction.

The energy spectrum of a particle in a $3D$ isotropic harmonic potential is%
\begin{equation}
E_{n}=n\hbar\omega\label{En_HO}%
\end{equation}
with the degenerate degree $\left(  n+1\right)  \left(  n+2\right)  /2$, where
the zero-point energy has been suppressed. Then the heat kernel is%
\begin{equation}
K\left(  t\right)  =\sum_{n=0}^{\infty}\frac{1}{2}\left(  n+1\right)  \left(
n+2\right)  e^{-nt}=\frac{1}{\left(  1-e^{-t}\right)  ^{3}}=\sum_{k=0}%
^{\infty}C_{k}t^{k-3},\label{Kt_3HO}%
\end{equation}
where the coefficients are%
\begin{equation}
C_{0}=1,C_{1}=\frac{3}{2},C_{2}=1,\cdots\label{Ck}%
\end{equation}
Although the form of the heat kernel expansion is different from eq.
(\ref{HKexp}), we can apply the same procedure to discuss the BEC phase
transition in the harmonic potential.

Replacing eq. (\ref{HKexp}) by eq. (\ref{Kt_3HO}) and substituting it into the
grand potential eq. (\ref{Xi}) gives%

\begin{equation}
\ln\Xi=\sum_{\ell=1}^{\infty}\frac{1}{\ell}\sum_{k=0}^{\infty}C_{k}\left(
\ell\beta\hbar\omega\right)  ^{k-3}z^{\ell}=\sum_{k=0}^{\infty}C_{k}\left(
\beta\hbar\omega\right)  ^{k-3}g_{4-k}\left(  z\right)  .
\end{equation}
Then the number of particles is%
\begin{equation}
N=\sum_{k=0}^{\infty}C_{k}\left(  \beta\hbar\omega\right)  ^{k-3}%
g_{3-k}\left(  z\right)  .\label{N_3HO1}%
\end{equation}
The critical temperature is given by eq. (\ref{N_3HO1}) at the limit
$\mu\rightarrow0$. However, at this limit, only the first two terms in the sum
are convergent. In the usual treatment, all of the divergent terms are ignored
\cite{KD,KT2,HHR,GH2}, and the result of the critical temperature only
contains the first-order correction. In the following, we will give the
second-order correction to the critical temperature.

Taking advantage of the asymptotic expression of the Bose-Einstein integral
eq. (\ref{BEint}), we have%
\begin{equation}
N\approx\frac{C_{0}}{\left(  \beta\hbar\omega\right)  ^{3}}\zeta\left(
3\right)  +\frac{C_{1}}{\left(  \beta\hbar\omega\right)  ^{2}}\zeta\left(
2\right)  -\frac{C_{2}}{\beta\hbar\omega}\ln\left(  -\beta\mu\right)
+I_{3},\label{N_3HO2}%
\end{equation}
where we have introduced%
\begin{equation}
I_{3}=\sum_{k=3}^{\infty}C_{k}\left(  \beta\hbar\omega\right)  ^{k-3}%
\Gamma\left(  k-2\right)  \frac{1}{\left(  -\beta\mu\right)  ^{k-2}}.
\end{equation}
After introducing the integral form of the gamma function with the
regularization parameter $s\rightarrow0$, eq. (\ref{Gamma}), we have%
\begin{align}
I_{3}  &  =\frac{1}{\left(  -\beta\mu\right)  }\int_{0}^{\infty}dxx^{s}%
e^{-x}\sum_{k=3}^{\infty}C_{k}\left(  \frac{\beta\hbar\omega}{-\beta\mu
}x\right)  ^{k-3}\nonumber\\
&  =\frac{1}{\left(  -\beta\mu\right)  }\int_{0}^{\infty}dxx^{s}e^{-x}K\left(
\frac{\hbar^{2}}{2m}\frac{x}{-\mu}\right) \nonumber\\
&  -\frac{C_{0}\Gamma\left(  s-2\right)  }{\beta\hbar\omega}\left(  \frac
{-\mu}{\hbar\omega}\right)  ^{2}-\frac{C_{1}\Gamma\left(  s-1\right)  }%
{\beta\hbar\omega}\frac{-\mu}{\hbar\omega}-\frac{C_{2}\Gamma\left(  s\right)
}{\beta\hbar\omega},\label{I3-2}%
\end{align}
where we have used the heat kernel eq. (\ref{Kt_3HO}) to perform the divergent
sum. The second and third terms in eq. (\ref{I3-2}) contain positive powers of
$\mu$, so they will vanish when $\mu\rightarrow0$. The first term can be
changed to a spectrum zeta function, i.e., a sum of the power of the spectrum,
by eq. (\ref{HK2sum}). Therefore, when $\mu\rightarrow0$, we have%
\begin{equation}
I_{3}=\Gamma\left(  1+s\right)  \frac{\left(  -\mu\right)  ^{s}}{\beta}%
\sum_{i=1}^{\infty}\frac{1}{E_{i}^{1+s}}-\frac{C_{2}}{\beta\hbar\omega}%
\Gamma\left(  s\right)  .\label{I3-3}%
\end{equation}
The sum of the excited-state spectrum can be performed exactly as%
\begin{equation}
\sum_{i=1}^{\infty}\frac{1}{E_{i}^{1+s}}=\frac{1}{2\left(  \hbar\omega\right)
^{1+s}}\left[  \zeta\left(  s-1\right)  +3\zeta\left(  s\right)
+2\zeta\left(  s+1\right)  \right]  .
\end{equation}
When $s\rightarrow0$, eq. (\ref{I3-3}) becomes%
\begin{equation}
I_{3}=\frac{1}{\beta\hbar\omega}\left(  -\ln\frac{\hbar\omega}{-\mu}-\frac
{19}{24}+\gamma_{E}\right)  ,\label{I3-4}%
\end{equation}
where the coefficients eq. (\ref{Ck}) have been used. In this result, the
linearly divergent terms of $s$ have been canceled. Substituting eqs.
(\ref{I3-4}) and (\ref{Ck}) into eq. (\ref{N_3HO2}) gives%
\begin{equation}
N=\frac{1}{\left(  \beta\hbar\omega\right)  ^{3}}\zeta\left(  3\right)
+\frac{3}{2}\frac{1}{\left(  \beta\hbar\omega\right)  ^{2}}\zeta\left(
2\right)  +\frac{1}{\beta\hbar\omega}\left[  -\ln\left(  \beta\hbar
\omega\right)  -\frac{19}{24}+\gamma_{E}\right]  .
\end{equation}
Just like the general case discussed in section \ref{II}, the logarithmically
divergent terms of $\mu$ are also canceled. This is a fully analytical result
of the critical temperature. Since $\beta\hbar\omega\ll1$ at the transition
point, the last two terms correspond the first- and second-order corrections.
Then the critical temperature is approximately%
\begin{align}
T_{c}  &  \approx\frac{N^{1/3}}{\zeta^{1/3}\left(  3\right)  }\frac
{\hbar\omega}{k_{B}}\left\{  1-\frac{\zeta\left(  2\right)  }{2\zeta
^{2/3}\left(  3\right)  }\frac{1}{N^{1/3}}+\frac{1}{9\zeta^{1/3}\left(
3\right)  }\left[  -\ln\frac{N}{\zeta\left(  3\right)  }+\frac{\pi
^{4}+38-48\gamma_{E}}{16\zeta\left(  3\right)  }\right]  \frac{1}{N^{2/3}%
}\right\} \nonumber\\
&  =\frac{N^{1/3}}{\zeta^{1/3}\left(  3\right)  }\frac{\hbar\omega}{k_{B}%
}\left[  1-0.7275\frac{1}{N^{1/3}}-\left(  0.1045\ln N-0.6045\right)  \frac
{1}{N^{2/3}}\right]  .\label{Tc_3HO}%
\end{align}
Compared with Ref. \cite{KD,KT2,HHR,GH2}, the first-order correction in eq.
(\ref{Tc_3HO}) is consistent with their results, and we also give the
second-order correction.

\subsection{Two-dimensional harmonic potentials}

After suppressing the zero-point energy, the spectrum of a particle in a
two-dimensional ($2D$) isotropic harmonic potential is still given by eq.
(\ref{En_HO}), but the degeneracy is $n+1$, so the heat kernel becomes%
\begin{equation}
K\left(  t\right)  =\sum_{n=0}^{\infty}\left(  n+1\right)  e^{-nt}%
=\frac{e^{2t}}{\left(  e^{t}-1\right)  ^{2}}=\sum_{k=0}^{\infty}C_{k}t^{k-2},
\end{equation}
where the expansion coefficients are%
\begin{equation}
C_{0}=1,C_{1}=1,C_{2}=\frac{5}{12},\cdots\label{Ck2}%
\end{equation}
Then the grand potential eq. (\ref{Xi}) and the number of particles become%
\begin{align}
\ln\Xi &  =\sum_{l=1}^{\infty}\frac{1}{l}\sum_{k=0}^{\infty}C_{k}\left(
l\beta\hbar\omega\right)  ^{k-2}z^{l}=\sum_{k=0}^{\infty}C_{k}\left(
\beta\hbar\omega\right)  ^{k-2}g_{3-k}\left(  z\right)  ,\nonumber\\
N  &  =\sum_{k=0}^{\infty}C_{k}\left(  \beta\hbar\omega\right)  ^{k-2}%
g_{2-k}\left(  z\right)  .
\end{align}
At the limit $\mu\rightarrow0$, by using the asymptotic form eq.
(\ref{BEint}), we can rewrite the expression of the particle number as
\begin{equation}
N\approx\frac{C_{0}}{\left(  \beta\hbar\omega\right)  ^{2}}\zeta\left(
2\right)  -\frac{C_{1}}{\beta\hbar\omega}\ln\left(  -\beta\mu\right)
+I_{2},\label{N_2HO}%
\end{equation}
where we have introduced%
\begin{equation}
I_{2}=\sum_{k=2}^{\infty}C_{k}\left(  \beta\hbar\omega\right)  ^{k-2}%
\Gamma\left(  k-1\right)  \frac{1}{\left(  -\beta\mu\right)  ^{k-1}}.
\end{equation}
Clearly, different from the $3D$ case, the second term in eq. (\ref{N_2HO}) is
divergent. By using the integral form of the gamma function eq. (\ref{Gamma}),
we obtain
\begin{align}
I_{2}  &  =\frac{1}{\left(  -\beta\mu\right)  }\int_{0}^{\infty}dxx^{s}%
e^{-x}\sum_{k=2}^{\infty}C_{k}\left(  \frac{x}{-\mu}\hbar\omega\right)
^{k-2}\nonumber\\
&  =\frac{1}{\left(  -\beta\mu\right)  }\int_{0}^{\infty}dxx^{s}e^{-x}K\left(
\frac{\hbar^{2}}{2m}\frac{x}{-\mu}\right)  -\frac{C_{0}\Gamma\left(
s-1\right)  }{\beta\hbar\omega}\frac{-\mu}{\hbar\omega}-\frac{C_{1}%
\Gamma\left(  s\right)  }{\beta\hbar\omega}.
\end{align}
When $\mu\rightarrow0$, the second term vanishes, and the first term becomes a
sum of the spectrum,%
\begin{equation}
I_{2}=\Gamma\left(  1+s\right)  \frac{\left(  -\mu\right)  ^{s}}{\beta}%
\sum_{i=1}^{\infty}\frac{1}{E_{i}^{1+s}}-\frac{1}{\beta\hbar\omega}%
\Gamma\left(  s\right)  ,
\end{equation}
where the coefficients eq. (\ref{Ck2}) have been substituted. The exact result
of the sum of the excited-state spectrum is%
\begin{equation}
\sum_{i=1}^{\infty}\frac{1}{E_{i}^{1+s}}=\frac{1}{\left(  \hbar\omega\right)
^{1+s}}\left[  \zeta\left(  s\right)  +\zeta\left(  1+s\right)  \right]  .
\end{equation}
Then taking $s\rightarrow0$ gives%
\begin{equation}
I_{2}=\frac{1}{\beta\hbar\omega}\left(  -\ln\frac{\hbar\omega}{-\mu}-\frac
{1}{2}+\gamma_{E}\right)  .
\end{equation}
The particle number eq. (\ref{N_2HO}) becomes%
\begin{equation}
N=\frac{1}{\left(  \beta\hbar\omega\right)  ^{2}}\zeta\left(  2\right)
+\frac{1}{\beta\hbar\omega}\left(  -\ln\left(  \beta\hbar\omega\right)
-\frac{1}{2}+\gamma_{E}\right)  .
\end{equation}
Just as before, all of the divergent terms have been canceled, so the critical
temperature of BEC in the $2D$ harmonic potential is%
\begin{align}
T_{c}  &  \approx\sqrt{\frac{N}{\zeta\left(  2\right)  }}\frac{\hbar\omega
}{k_{B}}\left[  1-\frac{1}{4\zeta^{1/2}\left(  2\right)  }\left(  \ln\frac
{N}{\zeta\left(  2\right)  }-1+2\gamma_{E}\right)  \frac{1}{N^{1/2}}\right]
\nonumber\\
&  =\sqrt{\frac{N}{\zeta\left(  2\right)  }}\frac{\hbar\omega}{k_{B}}\left[
1-\left(  0.1949\ln N-0.0670\right)  \frac{1}{N^{1/2}}\right] \label{Tc_2HO}%
\end{align}

\section{The second critical temperature of BEC \label{V}}

As is well known, when the BEC phase transition occurs, a large number of
particles will fall into the ground state and the ground state will be
macroscopically occupied. However, in some anisotropic systems, the behavior
of the Bose gas may be more complicated. If there exists a set of quantum
states which are very closed to the ground state, a macroscopic number of
particles may distribute over the set of states. This phenomenon is called the
generalized BEC \cite{BL,BLP,BLL}, which can be classified into three types
\cite{BL,BLP,Beau}: Type I (II) refers to the case that a finite (infinite)
number of single-particle states are macroscopically occupied; type III refers
to the case that the occupation of the set of the states is a macroscopic
fraction of the total particle number although none of these states is
macroscopically occupied. The generalized BEC has been discussed in various
geometries and external potentials, with and without interaction
\cite{BZ2,BZ3,BZ}.

In ref. \cite{BZ}, the authors discuss the generalized BEC in some anisotropic
systems\ in the thermodynamic limit, including slabs, squared beams, and
"cigars". They show that in these systems, ideal Bose gases will undergo two
kinds of phase transitions. Therefore, besides the conventional critical
temperature $T_{c}$, there is a second critical temperature $T_{m}$ connected
to the second phase transition. In this section, we will take the anisotropic
slab as an example to discuss the same problem without the assumption of the
thermodynamic limit. and we will give the correction of the boundary on the
second critical temperature of BEC.

We consider a Bose gas in a highly anisotropic slab described in section
\ref{slab}. As in ref. \cite{BZ}, the side length of the slab takes the form
$a=Le^{\alpha L}\gg L$. The total particle number can be expressed as%

\begin{align}
N &  =N_{0}\left(  T\right)  +N_{1}\left(  T\right)  +N_{2}\left(  T\right)
\nonumber\\
&  =f\left(  1,1,1\right)  +\sum_{n_{x},n_{y}\neq\left(  1,1\right)  }%
^{\infty}f\left(  n_{x},n_{y},1\right)  +\sum_{n_{z}=2}^{\infty}\sum
_{n_{x},n_{y}=1}^{\infty}f\left(  n_{x},n_{y},n_{z}\right)  ,\label{N_3}%
\end{align}
where $f\left(  n_{x},n_{y},n_{z}\right)  =\left(  z^{-1}e^{-\beta E\left(
n_{x},n_{y},n_{z}\right)  }-1\right)  ^{-1}$ denotes the average particle
number in the state $\left(  n_{x},n_{y},n_{z}\right)  $. In eq. (\ref{N_3})
we have divided the total particle number into three parts: the ground-state
particle number $N_{0}\left(  T\right)  $, the number of particles in the
states with $n_{z}=1$ (not including the ground state) $N_{1}\left(  T\right)
$, and the number of particles in the state with $n_{z}\geq2$, $N_{2}\left(
T\right)  $. The condition for the conventional BEC is then \cite{BZ}%
\begin{equation}
N=N_{2}^{\max}\left(  T_{c}\right)  .\label{N_Tc}%
\end{equation}
Since at the transition point $T_{c}$, the number of the particles in the
states with $n_{z}=1$ is negligible, $N_{2}\left(  T_{c}\right)  $ can be
replaced by the total number of the excited-state particles. It means that the
condition eq. (\ref{N_Tc}) is actually equivalent to the condition for the
conventional BEC. The critical temperature $T_{c}$ has been given in section
\ref{slab}, which satisfies eq. (\ref{n_slab}), or,%
\begin{equation}
N=\frac{V}{\lambda_{c}^{3}}\zeta\left(  \frac{3}{2}\right)  +\frac{a^{2}%
}{2\lambda_{c}^{2}}\ln\left(  \frac{\lambda_{c}^{2}}{16\pi L^{2}}\right)
.\label{N_slab}%
\end{equation}

Different from the usual three-dimensional system, in the highly anisotropic
slab, when $T<T_{c}$, no single state is macroscopically occupied, but the
entire band of the states with $n_{z}=1$ is macroscopically occupied, i.e.,
this is a type III generalized condensation. Only when the temperature is
lower than the second critical temperature, $T<T_{m}$, the ground state
becomes macroscopically occupied. Then there is a coexistence of the type III
generalized condensation and the standard type I condensation in the ground
state \cite{BZ}. As a result, the second critical temperature $T_{m}$ satisfies%

\begin{equation}
N=N_{1}^{\max}\left(  T_{m}\right)  +N_{2}^{\max}\left(  T_{m}\right)
.\label{N_Tm}%
\end{equation}
In the following we will discuss the influence of the boundary to the second
critical temperature.

Since%

\begin{equation}
N_{1}\left(  T\right)  =\sum_{n_{x},n_{y}=1}^{\infty}\frac{1}{z^{-1}e^{\beta
E\left(  n_{x},n_{y},1\right)  }-1}-\frac{1}{z^{-1}e^{\beta E\left(
1,1,1\right)  }-1},
\end{equation}
when $a\rightarrow\infty$, the sums of $n_{x},n_{y}$ can be converted to
integrals, we have%
\begin{equation}
N_{1}\left(  T\right)  =\frac{a^{2}}{\lambda^{2}}g_{1}\left(  e^{-\beta\Delta
}\right)  -\frac{1}{e^{\beta\Delta}-1},
\end{equation}
where we have introduced%
\begin{equation}
\Delta=\frac{\pi^{2}\hbar^{2}}{2mL^{2}}-\mu.
\end{equation}
Nearby the second critical temperature, $\beta\Delta\rightarrow0$, then%
\begin{equation}
N_{1}\left(  T\right)  \approx-\frac{a^{2}}{\lambda^{2}}\ln\left(  \beta
\Delta\right)  -\frac{1}{\beta\Delta}.
\end{equation}
Since its maximum appears at $\beta\Delta=\lambda_{m}^{2}/a^{2}$, the
condition for the phase transition eq. (\ref{N_Tm}) becomes%
\begin{equation}
N\approx\frac{a^{2}}{\lambda_{m}^{2}}\ln\left(  \frac{a^{2}}{4\sqrt{\pi
}L\lambda_{m}}\right)  -\frac{a^{2}}{\lambda_{m}^{2}}+\zeta\left(  \frac{3}%
{2}\right)  \frac{V}{\lambda_{m}^{3}}.
\end{equation}
Substituting $a=Le^{\alpha L}$ gives%
\begin{equation}
N=\frac{L^{2}e^{2\alpha L}}{\lambda_{m}^{2}}\left[  2\alpha L-1+\ln\left(
\frac{L}{4\sqrt{\pi}\lambda_{m}}\right)  \right]  +\zeta\left(  \frac{3}%
{2}\right)  \frac{L^{3}e^{2\alpha L}}{\lambda_{m}^{3}}.\label{N_2nd}%
\end{equation}
This is the equation that the second critical temperature $T_{m}$ obeyed.
Taking advantage into eq. (\ref{N_slab}), we can obtain the relation between
the two critical temperatures%
\begin{equation}
T_{m}^{3/2}+\frac{\xi}{L}T_{m}\left[  2\alpha L-1+\frac{1}{2}\ln\left(
\frac{L^{2}mk}{32\pi^{2}\hbar^{2}}T_{m}\right)  \right]  =T_{c}^{3/2}%
-\frac{\xi}{2L}T_{c}\ln\left(  \frac{L^{2}mk}{8\hbar^{2}}T_{c}\right)
,\label{Tm&Tc}%
\end{equation}
where%
\begin{equation}
\xi=\frac{1}{\zeta\left(  3/2\right)  }\sqrt{\frac{2\pi\hbar^{2}}{mk}}.
\end{equation}
Easy to check that in the thermodynamic limit $L\rightarrow\infty$, the
relation (\ref{Tm&Tc}) becomes%
\begin{equation}
T_{m}^{3/2}+2\alpha\xi T_{m}=T_{c}^{3/2},
\end{equation}
which is consistence with the result given in ref. \cite{BZ}.

On the other hand, if we consider an isothermal process, the conditions for
the phase transitions will described by the critical particle densities.
According to eqs. (\ref{N_slab}) and (\ref{N_2nd}), in an isothermal process,
the two critical densities are%

\begin{align}
n_{c} &  =\frac{\zeta\left(  3/2\right)  }{\lambda^{3}}+\frac{1}{L\lambda^{2}%
}\ln\left(  \frac{\lambda}{4\sqrt{\pi}L}\right)  ,\nonumber\\
n_{m} &  =\frac{\zeta\left(  3/2\right)  }{\lambda^{3}}+\frac{1}{L\lambda^{2}%
}\left[  2\alpha L+\ln\left(  \frac{L}{4\sqrt{\pi}\lambda}\right)  -1\right]
.
\end{align}
Then we have%
\begin{equation}
n_{m}=n_{c}+\frac{2\alpha}{\lambda^{2}}+\frac{2}{L\lambda^{2}}\left(  \ln
\frac{L}{\lambda}-\frac{1}{2}\right)  .
\end{equation}
In the thermodynamic limit $L\rightarrow\infty$, this relation will go back to
the result in ref. \cite{BZ}.

In the above we have obtained the correction of the boundary to the second
critical temperature of BEC in a slab geometry. The similar method can also be
applied to the systems with other anisotropic boundaries or external potentials.

\section{Conclusion and discussions \label{VI}}

In the above, we discuss the critical temperature of BEC of ideal gases in
finite systems in a general framework based on the heat kernel expansion and
the zeta-function regularization. Our method gives the analytical expression
of the critical temperature only related to the heat kernel coefficients. We
consider some specific examples and give the corresponding critical
temperatures. Some of them have been obtained by other methods, but we provide
a consistent treatment for different systems in this paper. Besides, taking
advantage of the asymptotic spectrum, we divide the effect of the finite
number of particles on the critical temperature from other factors, and the
result agrees with the numerical calculation very well. For the Bose gas in
isotropic harmonic traps, we obtain the second-order correction to the
critical temperature for $3D$ case and the first-order correction for $2D$
case. In some highly anisotropic systems, besides the conventional BEC, the
generalized condensation may occur in a Bose system. We also give the
correction of the boundary on the second critical temperature in an
anisotropic slab. We hope that our work can help to reveal the nature of the
phase transition in finite systems.

In this paper, we only discuss the critical temperature of BEC. Besides these
results, our method can also be applied to other aspects of the phase
transition. For example, it can be used to analyze the properties of the
thermodynamic functions, especially near the transition point. We will leave
these for future work.

\bigskip

The author is very indebted to Prof. Wu-Sheng Dai for his help. This work is
supported in part by NSF of China, under Project No. 11575125.

\end{document}